\pdfoutput=1

\documentclass[conference]{IEEEtran}
\usepackage[utf8]{inputenc}

\usepackage{xcolor}
\usepackage{listings}

\newcommand\JSONnumbervaluestyle{\color{blue}}
\newcommand\JSONstringvaluestyle{\color{red}}

\newif\ifcolonfoundonthisline

\makeatletter

\lstdefinestyle{json}
{
  showstringspaces    = false,
  keywords            = {false,true},
  alsoletter          = 0123456789.,
  morestring          = [s]{"}{"},
  stringstyle         = \ifcolonfoundonthisline\JSONstringvaluestyle\fi,
  MoreSelectCharTable =%
    \lst@DefSaveDef{`:}\colon@json{\processColon@json},
  basicstyle          = \ttfamily,
  keywordstyle        = \ttfamily\bfseries,
}

\newcommand\processColon@json{%
  \colon@json%
  \ifnum\lst@mode=\lst@Pmode%
    \global\colonfoundonthislinetrue%
  \fi
}

\lst@AddToHook{Output}{%
  \ifcolonfoundonthisline%
    \ifnum\lst@mode=\lst@Pmode%
      \def\lst@thestyle{\JSONnumbervaluestyle}%
    \fi
  \fi
  \lsthk@DetectKeywords%
}

\lst@AddToHook{EOL}%
  {\global\colonfoundonthislinefalse}

\usepackage{cite}

\usepackage{tabularx}

\ifCLASSINFOpdf
  \usepackage[pdftex]{graphicx}
  \graphicspath{{../pdf/}{../jpeg/}{./fig}}
  \DeclareGraphicsExtensions{.pdf,.jpeg,.png,.jpg}
\else
\fi
\usepackage{amsmath}
\usepackage{array}

\ifCLASSOPTIONcompsoc
  \usepackage[caption=false,font=normalsize,labelfont=sf,textfont=sf]{subfig}
\else
  \usepackage[caption=false,font=footnotesize]{subfig}
\fi

\usepackage{xspace}
\usepackage{xcolor,color}
\usepackage{blkarray}
\usepackage{multirow}
\usepackage{subfig}
\usepackage{graphicx}
\usepackage{epstopdf}
\usepackage{wrapfig}

\newcommand{\ourname}{\textsc{IoT Sentinel}\xspace}
\newcommand{\securebox}{Security Gateway\xspace}
\newcommand{\iotssp}{IoTSSP\xspace}
\newcommand{\securityservice}{IoT Security Service\xspace}

\begin{document}
%
\title{\ourname: Automated Device-Type Identification for Security Enforcement in IoT}

\author{\IEEEauthorblockN{Markus Miettinen\\Ahmad-Reza Sadeghi}
\IEEEauthorblockA{Technische Universität Darmstadt, Germany\\
Email: \{markus.miettinen,ahmad.sadeghi\}\\@trust.tu-darmstadt.de}
\and
\IEEEauthorblockN{Samuel Marchal\\N. Asokan}
\IEEEauthorblockA{Aalto University, Espoo, Finland\\
Email: samuel.marchal@aalto.fi\\asokan@acm.org}
\and
\IEEEauthorblockN{Ibbad Hafeez\\Sasu Tarkoma}
\IEEEauthorblockA{University of Helsinki, Finland\\
Email: \{ibbad.hafeez,sasu.tarkoma\}\\@cs.helsinki.fi}}


%


%
\maketitle

\begin{abstract}
With the rapid growth of the Internet-of-Things (IoT), concerns about the security of IoT devices have become prominent. Several vendors are producing IP-connected devices for home and small office networks that often suffer from flawed security designs and implementations. They also tend to lack mechanisms for firmware updates or patches that can help eliminate security vulnerabilities. Securing networks where the presence of such vulnerable devices is given, requires a \emph{brownfield approach}: applying necessary protection measures within the network so that potentially vulnerable devices can coexist without endangering the security of other devices in the same network.
In this paper, we present \ourname, a system capable of automatically identifying the types of devices being connected to an IoT network and enabling enforcement of rules for constraining the communications of vulnerable devices so as to minimize damage resulting from their compromise. We show that \ourname is effective in identifying device types and has minimal performance overhead.
\end{abstract}

%

\section{Introduction}
\label{sect:intro}

The proliferation of the Internet-of-Things (IoT) is an ongoing megatrend in computing with recent forecasts suggesting the number of IoT devices to reach 24 billion in 2020~\cite{Businessinsider2016}. More and more people install IP-enabled devices and household appliances in their homes in order to benefit from the improved ability to be informed about and control relevant features of their homes. Examples of emerging IoT systems include automated heating and air conditioning, security systems and home surveillance, lighting or traditional household appliances with added WiFi-connectivity. Also entirely new device classes utilizing network connectivity are emerging.

Numerous device vendors are providing such connected products to users. Many of these firms are traditional household appliance manufacturers and do not necessarily have expertise in engineering systems with computer security in mind. As a result, there are many reports in the media about IoT devices being deployed in users' homes with security vulnerabilities that can be exploited by attackers (e.g., \cite{Core2013,Pentestpartners2015}). 
There have been reports of a single software flaw affecting a full range of different products, as software components are reused for different device models, thereby placing thousands of Internet-connected IoT devices susceptible for attack~\cite{Senrio2016}.
Recent reports also suggest that a considerable number of deployed devices use publicly known private keys~\cite{SECConsult2016}, thus making them vulnerable to unauthorized access by external adversaries. 

Using vulnerabilities in insecure devices, adversaries can mount attacks against the user's home network. The preferred solution for dealing with device vulnerabilities would be to patch them in order to eliminate weaknesses. However, all too often device vendors are either unable (many users do not register their devices with the device vendor) or unwilling to provide such patches in a timely manner.
Most IoT users do not have the skills or willingness to perform such tasks or they even forget unattended IoT devices previously installed in their network leaving them with outdated software versions.

Future security solutions for IoT will need to take into account that IoT devices with unpatched vulnerabilities may often be present in the user's network and co-exist with other devices during their whole device lifetime. 
The presence of insecure, unpatched legacy IoT devices mandates to accommodate a \emph{brownfield} development approach for security designs: the security mechanisms must be able to co-exist with potentially insecure devices and software that users already have deployed or will deploy in their home networks.

\textbf{Goals and Contributions.} In this paper, we tackle this problem by presenting  \ourname, a system capable of identifying the types of devices introduced to a network and enforcing mitigation measures for device-types that have potential security vulnerabilities.
\ourname does so by controlling the traffic flows of vulnerable devices in order to protect other devices in the network from threats and prevent data leakage.


The contributions of this paper are the following:
\begin{itemize}
	\item We present the design of \ourname, a security system (Sect.~\ref{sect:system}) for managing the security and privacy risks posed by inherently insecure IoT devices deployed in users' networks.
	\item We introduce a device-type identification technique tailored for IP-enabled IoT devices (Sect.~\ref{sect:identification}).
	Device-type identification, in conjunction with information from vulnerability databases can pinpoint vulnerable devices in a network.
	\item  We demonstrate the accuracy and scalability of \ourname device-type identification using a large set of different real-world off-the-shelf IoT devices (Sect.~\ref{sect:evaluation}).
	\item We present a framework for confining traffic flows of devices identified as vulnerable. The framework uses software-defined networking to implement network isolation and traffic filtering. We also show that the enforcement framework can be efficiently implemented with moderate impact on traffic latency (Sect.~\ref{sect:enforcement}).

\end{itemize}

\section{Adversary Model}
\label{sect:adversary}
\ourname is targeted at a typical network setup found in homes and small offices, where devices are connected to a gateway router offering wireless and wired interfaces for connecting IP-enabled devices to the network. 
We assume that when IoT devices are initially connected to the target network they possibly have security vulnerabilities but are initially benign, i.e., uncompromised by the adversary.
The adversary's goal is to exploit IoT devices to either a) exfiltrate data, security credentials or encryption keys, b) compromise other IoT devices in the network with the help of a compromised device, or, c) inject false or tampered information into the user's network.




It has been demonstrated~\cite{Sivaraman2016} that also remote attacks against IoT devices are feasible, if devices like smartphones that are infected with malware are used to locate vulnerable IoT devices inside the user's WiFi network and to open an attack path for the remote attacker through ``NAT hole punching'', allowing the attacker to connect to the target device remotely for executing the attack. This emphasizes the importance of the gateway router for upholding the security of the IoT network.


The goal of \ourname is to restrict communications in the network so that the adversary is either not able to connect to the vulnerable device to exploit vulnerabilities, unable to use a compromised device to attack other devices in the network, or, unable to exfiltrate data from compromised devices, thereby effectively mitigating attacks or limiting their impact.

\section{System Design}
\label{sect:system}
\begin{figure}
	\centering
	\includegraphics[width=.8\columnwidth]{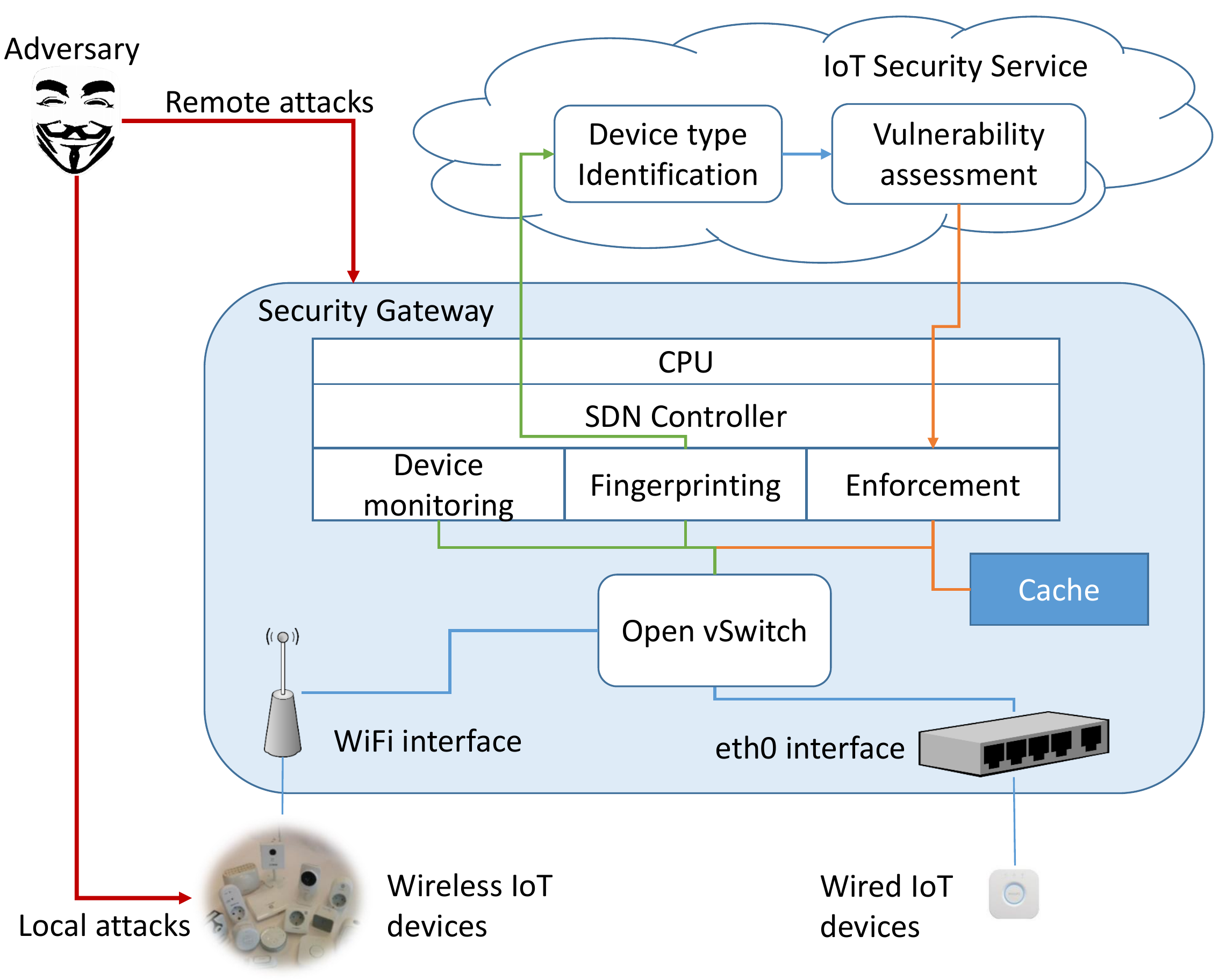}
	\caption{\ourname system design}
	\label{fig:system-design}
\end{figure}

To protect the network against adversaries, \ourname will 1) identify the device-types of new IoT devices introduced into the network, 2) make a vulnerability assessment of a device using its device-type, and, 
3) constrain communication capabilities of the device accordingly. 
In this paper we focus on \#1 and \#3. 
The term \emph{device-type} in this work is defined to denote the combination of \emph{make}, \emph{model} and \emph{software version} of a particular device. 
Device-type identification in \ourname is based on monitoring the communication behaviour of devices during the setup process to generate device-specific fingerprints which are mapped to device-types with the help of a machine learning-based classification model. 
For a given device-type, its potential vulnerability can be assessed by consulting an external information source as we briefly describe in Sect.~\ref{sect:securityservice}.
Based on the vulnerability assessment, \ourname protects the target network by limiting network communications of the vulnerable device accordingly.
The design of our solution, shown in Fig.~\ref{fig:system-design}, consists of two major components: a \emph{\securebox} located in the user's local network and an \emph{\securityservice} operated by an IoT Security Service Provider (\iotssp). 

\subsection{\securebox}
The \securebox is a software-defined networking (SDN) based traffic monitoring and control component acting as a gateway router of a home or small office. Devices in the local network connect to the \securebox either through WiFi or an Ethernet connection. We envisage that \securebox will be deployed either as a dedicated hardware device, or, as a software module or firmware update for legacy WiFi access points having sufficient computational resources for running the \securebox functionality.

Wireless devices use WiFi Protected Setup (WPS) to obtain device-specific credentials in the form of WPA2 Pre-Shared Keys (PSK) for authenticating with the wireless interface of the \securebox.
This limits a local adversary's ability to impersonate other devices and eavesdrop on encrypted WiFi traffic in case the adversary can successfully compromise a device, as each device has a unique, device-specific PSK. 
For devices that do not support WPS, \securebox will provide a device-specific WPA2-PSK to be used in the setup process for WiFi authentication.

The \securebox monitors and profiles the behaviour of individual devices and sends \emph{device fingerprints} to the \securityservice for identification (Sect.~\ref{sect:classification}) and vulnerability assessment of individual devices. Based on this assessment, the \securityservice returns an \emph{isolation level} (Sect.~\ref{sect:enforcement}) to be enforced by the \securebox on the device. 

\subsection{\securityservice}
\label{sect:securityservice}

Based on device fingerprints provided by {\securebox}s, the \iotssp uses machine learning-based classification models to classify devices according to their \emph{device-type}.
Initially, ground truth information about device-types for training the models needs to be gathered manually by dedicated laboratory experiments with IoT devices. Crowdsourcing approaches could also be considered, in a similar way as anti-virus vendors mutually share malware signatures to improve detection accuracy.
Contrary to traditional fingerprinting approaches utilising distinctive characteristics of particular data fields in protocol messages, our approach focuses on the behavioural characteristics of devices as discussed in detail in Sect.~\ref{sect:identification}. 
This allows us to profile and classify devices without prior knowledge about the syntax of messages or data field values used by individual devices.

For each device-type in the training data, the \iotssp performs a vulnerability assessment. 
This assessment is based on querying  repositories like the CVE database~\cite{CVE2016} for vulnerability reports related to the device-type. The vulnerability assessment can be augmented with automated penetration tests or recently developped automated bug searches techniques for IoT devices~\cite{Feng:2016:scalable}. Crowdsourced information can also be used by cross-correlating security incidents and related device-types as reported by {\securebox}s of affected networks. In case vulnerabilities exist, isolation level \emph{restricted} (cf. Sect~\ref{sect:enforcement}) is assigned. If no vulnerabilities for the device-type are reported, it is assigned the level \emph{trusted}. Unknown devices will be assigned the level \emph{strict}.


Using the trained classifiers, the \iotssp identifies the device-types of any new devices by feeding their device fingerprints to the classifier.
Based on each device-type's vulnerability assessment, the \iotssp determines the appropriate isolation level required for the device and notifies the \securebox along with possible auxiliary information depending on the prescribed isolation level.
\securityservice does not store any information about its \securebox clients, it just receives fingerprints and returns an isolation level accordingly.
\securebox can anonymously request the \securityservice through anonymization networks such as Tor~\cite{dingledine2004tor} to ensure privacy preservation.


\subsection{Mitigation Strategies}
To manage vulnerable devices, we apply the following mitigation strategies that aim at maintaining as much functionality as possible while minimizing the risk of harm.
The technical implementation of mitigation measures is discussed in Sect.~\ref{sect:enforcement}.

\subsubsection{Network Isolation}
The target of network isolation is to block a potentially vulnerable IoT device from communicating with other devices so that it cannot mount attacks against them. To this end, the \securebox divides the user's network into two virtual network overlays: an untrusted and a trusted network.
Vulnerable devices are placed in the untrusted network and strictly isolated from other devices.


\subsubsection{Traffic Filtering}
The target of traffic filtering is to hinder adversaries from communicating with vulnerable devices and exploiting vulnerabilities or exfiltrating data. Traffic filtering is performed by the \securebox and can be
targeted at particular protocols or endpoints so that the functionality of the vulnerable device is affected as little as possible.

%

\subsubsection{User Notification}
In some cases, network isolation and traffic filtering are not sufficient to provide adequate protection,
e.g., if a vulnerable IoT device is equipped with an external communication channel like Bluetooth or an LTE data connection that cannot be controlled by the \securebox. Since a compromised device could use this channel for exfiltrating sensitive data, the only effective measure for securing the user's network is to manually remove devices at risk. 
We therefore envisage a mechanism by which the system notifies the user about devices with insurmountable security flaws, helps her to identify the device in question and make sure that it really is removed from the user's network.



\section{IoT Device Identification}
\label{sect:identification}

As we discuss in more detail in Sect.~\ref{sect:soa_fingerprinting},
existing device identification approaches based on wireless communication fingerprinting have drawbacks that limit their usability in our scenario. Some approaches focus on identifying particular hardware or driver-specific characteristics (e.g., \cite{Bratus:2008:active,cache2006fingerprinting,Franklin:2006:passive,maurice2013improving}), which is not sufficiently distinctive, as the same hardware components and drivers may be deployed in a wide variety of different device types.
Other approaches utilize unique hardware-specific characteristics like clock skew to identify unique network interface cards (e.g., \cite{Kohno2005,Jana:2008:fast,Arackaparambil:2010:reliability,Ureten2007,Brik:2008:wireless}). This is, however, too specific for our use, as we do not wish to identify individual device instances, but rather identify the \emph{device-type}, i.e., the combination of make, model and software version of each new device introduced into the target network.
The few techniques~\cite{Gao:2010:passive,Radhakrishnan2015} addressing device-type identification are designed for high-end devices generating a large amount of network traffic over a long period of time. However, smart home IoT devices generate few network traffic in a discontinuous manner, which makes these techniques inapplicable in our scenario.

In this section we therefore introduce fingerprints specifically designed to discriminate smart home device-types. 
A two-fold classification system fed with these fingerprints determines the type of unknown devices. The system is tailored for IoT scenarios being able to scale and adapt with minimal cost to a large and variable set of device-types. 
It is primarily aimed at devices communicating over WiFi but is also applicable to any communications channel supporting TCP/IP.
 
\subsection{Device Fingerprint}
\label{sect:fingerprinting}

Our fingerprint is based on passively observed network traffic. 
It leverages the specificity of smart home devices that need to be inducted into the home network and associated to the gateway by following a device/vendor specific procedure. This procedure is characterized by a distinguishable sequence of communications initiated by the inducted device, which our fingerprint attempts to capture.
When a new device identified by a newly observed MAC address starts communicating with the gateway, the latter records $n$ packets $\left\lbrace p_1, p_2, p_3, \hdots , p_n \right\rbrace$ received from it during its setup phase. 
The end of the setup phase can be automatically identified by a decrease in the rate of packets sent.
We extract 23 features, giving a vector representation for each packet  $p_i = \left\lbrace f_{1,i}, f_{2,i}, f_{3,i}, \hdots , f_{23,i} \right\rbrace$ where $ i \in\left\lbrace 1, \hdots ,n \right\rbrace $. Hence, a device fingerprint is a $23 \times n$ matrix $\mathbf{F}$ with each column representing a packet received with order $ i \in\left\lbrace 1, \hdots ,n \right\rbrace $ and each row representing a packet feature, see Eq.~\eqref{eq:feature_matrix}. Consecutive identical packets from our feature set perspective (i.e. $ p_i = p_{i+1}$) are discarded from $\mathbf{F}$.

\begin{equation}
  \mathbf{F} = 
    \begin{blockarray}{ccccc}
         p_1 & p_2 & p_3 & \hdots & p_n\\
      \begin{block}{(ccccc)}
        f_{1,1} & f_{1,2} & f_{1,3} & \hdots & f_{1,n} \\
        f_{2,1} & f_{2,2} & f_{2,2} & \hdots & f_{2,n} \\
        f_{3,1} & f_{3,2} & f_{3,3} & \hdots & f_{3,n} \\
        \vdots & \vdots & \vdots & \ddots & \vdots \\
        f_{23,1} &  f_{23,2}&  f_{23,3} & \hdots &  f_{23,n} \\
      \end{block}
    \end{blockarray}
\label{eq:feature_matrix}
\end{equation}

Features used for packet representation are presented in Table~\ref{tbl:pkt_features}, none of them rely on packet payload, ensuring that fingerprints can be extracted from encrypted traffic. A first set is composed of binary features set to $1$ if some selected communication protocols are used. These 16 protocols were chosen because they are typically used during device association over WiFi. Two binary features represent the use of IP header options \textit{padding} and \textit{router alert}. The size of the packet (in Bytes) and the presence of raw data is captured. The destination IP address, if any, is mapped to a counter starting from $1$ and incremented each time a new destination IP address is observed. This feature denotes the count and order in which a device communicates with different entities during its setup procedure. The two last features represent the source and destination ports used, if any, mapped to network port class:
\begin{itemize}
	\item no port $ \Rightarrow f = 0$
	\item \textit{well-known} port $\left[ 0 , 1023 \right] \Rightarrow f = 1$
	\item \textit{registered} port $\left[ 1024 , 49151 \right] \Rightarrow f = 2$
	\item \textit{dynamic} port $\left[ 49152 , 65535 \right] \Rightarrow f = 3$
\end{itemize}

\begin{table}[t]
\caption{Description of the 23 packet features. Features are binary except those marked with ``(int)'', which are integer.} \centering
	\begin{tabular}{l l }
	\textbf{Type}	& \textbf{Features}  \\ \hline
Link layer protocol (2) & ARP / LLC \\
Network layer protocol (4) & IP / ICMP / ICMPv6 / EAPoL \\
Transport layer protocol (2) &  TCP / UDP \\
\multirow{2}{*}{Application layer protocol (8)}  &  HTTP / HTTPS / DHCP / BOOTP / \\
&  SSDP / DNS / MDNS / NTP \\
IP options (2) &  \textit{Padding} / \textit{RouterAlert} \\
Packet content (2) & Size (int) / Raw data \\
IP address (1) & Destination IP counter (int) \\
Port class (2) &  Source (int) / Destination (int) \\

		\label{tbl:pkt_features}
 \end{tabular}
 \vspace*{-0.3cm}
\end{table}

Our fingerprints consider the temporal dimension of communication by capturing and keeping in its representation the sequential order in which packets are sent by a device $p_1 \rightarrow p_n$. In contrast to techniques aggregating network traffic statistics over a period of time \cite{Franklin:2006:passive,Pang:2007:user}, this extraction method raises some issues for comparison, since fingerprints have variable size $n$. To cope with this limitations we build a second fixed-size fingerprint $\mathbf{F'}$. 
$\mathbf{F'}$ is composed of the 12 first unique vector packets $p$ from $\mathbf{F}$ concatenated to produce a $276$-dimensional feature vector (12 packets $\times$ 23 features):\\
 $\mathbf{F'} = \left\lbrace f_{1,1}, f_{2,1}, \hdots, f_{23,1}, f_{1,2}, f_{2,2}, \hdots, f_{22,i}, f_{23,i}  \right\rbrace$

Preliminary analysis concluded that $12$ packets was a good trade-off for $\mathbf{F'}$ length: long enough to distinguish device-types and short enough to be fully filled with unique packets from $\mathbf{F}$.
However, if $\mathbf{F}$ does not contain enough unique packets to fill $\mathbf{F'}$, a padding with $0$ values is used to reach the size of 276 features. This produces a fixed size feature vector $\mathbf{F'}$ that can be used with standard machine learning techniques.

\subsection{Device-Type Identification}
\label{sect:classification}

In order to be scalable and applicable for an evolving number of devices, we propose an identification technique that is two-fold.
A first operation consists of building a single classifier for each device-type we have fingerprints for. Each classifier provides a binary decision whether the input fingerprint matches the device-type or not. An unknown fingerprint can be accepted by several classifiers and thus match several device-types.
In such cases, a second step discriminates the multiple matches using an edit distance based metric.
While edit distance could be used alone to identify device-types, this procedure is far more time consuming than classification as we will see in Sect~\ref{sect:identification_eval}. The classification step can easily apply to thousands of device-types, providing a limited set of device-types to discriminate via edit distance, ensuring the speed and scalability of the approach.

\subsubsection{Fingerprint  Classification}

The device classification is operated using the fixed length fingerprints $\mathbf{F'}$.
Let's assume we have a set of fingerprints $S$ for several device-types. We select the subset of $n$ fingerprints $S_{D_i} = \left\lbrace \mathbf{F'}_{1,i}, \mathbf{F'}_{2,i}, \hdots, \mathbf{F'}_{n,i}\right\rbrace$ for the device-type $D_i$. The remaining fingerprints of the set are for device-types $D_x \ne D_i$. These fingerprints belong to the complement of $S_{D_i}$ in $S$: $S^{c}_{D_i}$. 
A classifier $C_i$ is trained for identifying the device-type $D_i$, using all samples from $S_{D_i}$ as one class and a subset of samples from $S^{c}_{D_i}$ as the other class. 
Only a subset from $S^{c}_{D_i}$ is selected for classifier training in order to avoid imbalanced class learning issues \cite{He:2009:learning}.
$C_i$ is then able to identify an unknown fingerprint as belonging to the type $D_i$ or not.
This process is repeated for each device-type in $S$ in order to build one classifier per device-type.
We use Random Forest classification algorithm \cite{Breiman2001} to build these models.

Using this approach, every time the fingerprint of a new device-type is captured, a new classifier is trained without making any modification to the existing classifiers, avoiding a costly relearning process. This ``one classifier per device-type'' approach also enables the discovery of new devices since it does not force any fingerprint to belong to one learned class of a multi-class classifier. A fingerprint can be rejected by all classifiers and thus be identified as a new device-type.

\subsubsection{Edit Distance Discrimination}

If an unknown fingerprint $\mathbf{F'}$ matches several device-types during the classification process, the corresponding full fingerprint $\mathbf{F}$ is compared to a subset of fingerprints from each device-type it got a match for.
The fingerprint comparison is done by computing Damerau-Levenshtein edit distance \cite{Damerau:1964:technique} considering the insertion, deletion, substitution and immediate transposition of characters.
We consider the matrix $\mathbf{F}$ as a word with each character being a column of the matrix, i.e. a packet $p_i$. Character equality for edit distance computation is considered if all features $f$ from a packet $p_i$ are equal to those of another packet $p_j$. The obtained absolute distance between two fingerprints is divided by the length of the longest one to provide a normalized distance value bounded on $[0 , 1]$. 

The distance is computed between the fingerprint to identify $\mathbf{F}$ and a subset of five fingerprints from each device-type $D_i$ it got a match for. Distances are summed up per device-type to get a global dissimilarity score $s_i \in [0,5]$ of $\mathbf{F}$ with the type $D_i$. The lowest dissimilarity score $s_i$ gives the final predicted device-type for $\mathbf{F}$.

\section{Enforcement}
\label{sect:enforcement}
\securebox uses Software-defined Networking (SDN) to enable enforcement. We wrote a custom module for Floodlight SDN controller~\cite{floodlight-controller} to perform network monitoring tasks, fingerprint generation and to manage communications with \securityservice. This module is also responsible for generation and enforcement of restricted network access for connected devices.

\begin{figure}
	\includegraphics[width=0.9\columnwidth]{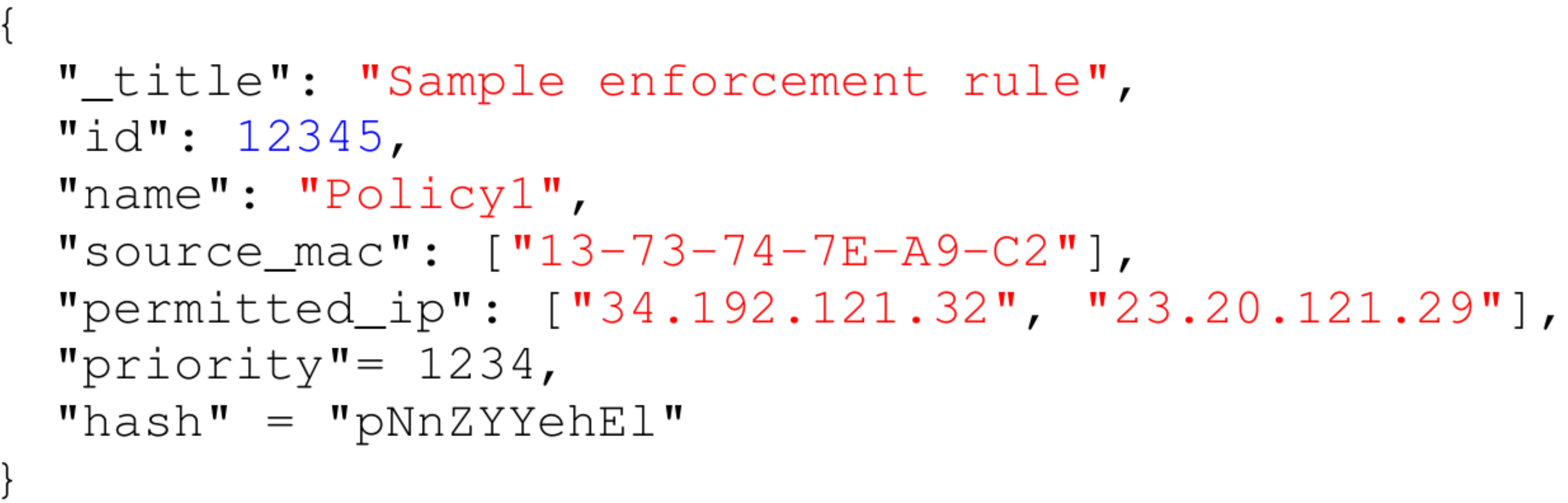}
	\caption{\textbf{Sample enforcement rule}. \textsl{Rules are specified for single devices using their MAC addresses. If the device isolation level is \textit{Restricted}, a list of \textit{permitted IP} addresses is given through which the device can communicate with its cloud service. The \textit{hash} value is used for enforcement rule storage in cache.}}
	\label{fig:enforcement-rule} 
\end{figure}

When a new device connects to the network, \securebox generates a fingerprint from its network activity. This fingerprint is sent to \securityservice, which identifies the device-type, determines its required network isolation level and returns it to the \securebox. There are three different isolation levels for any device as shown in Fig.~\ref{fig:isolation-levels}: 

\begin{figure}
	\centering
	\includegraphics[width=.8\columnwidth]{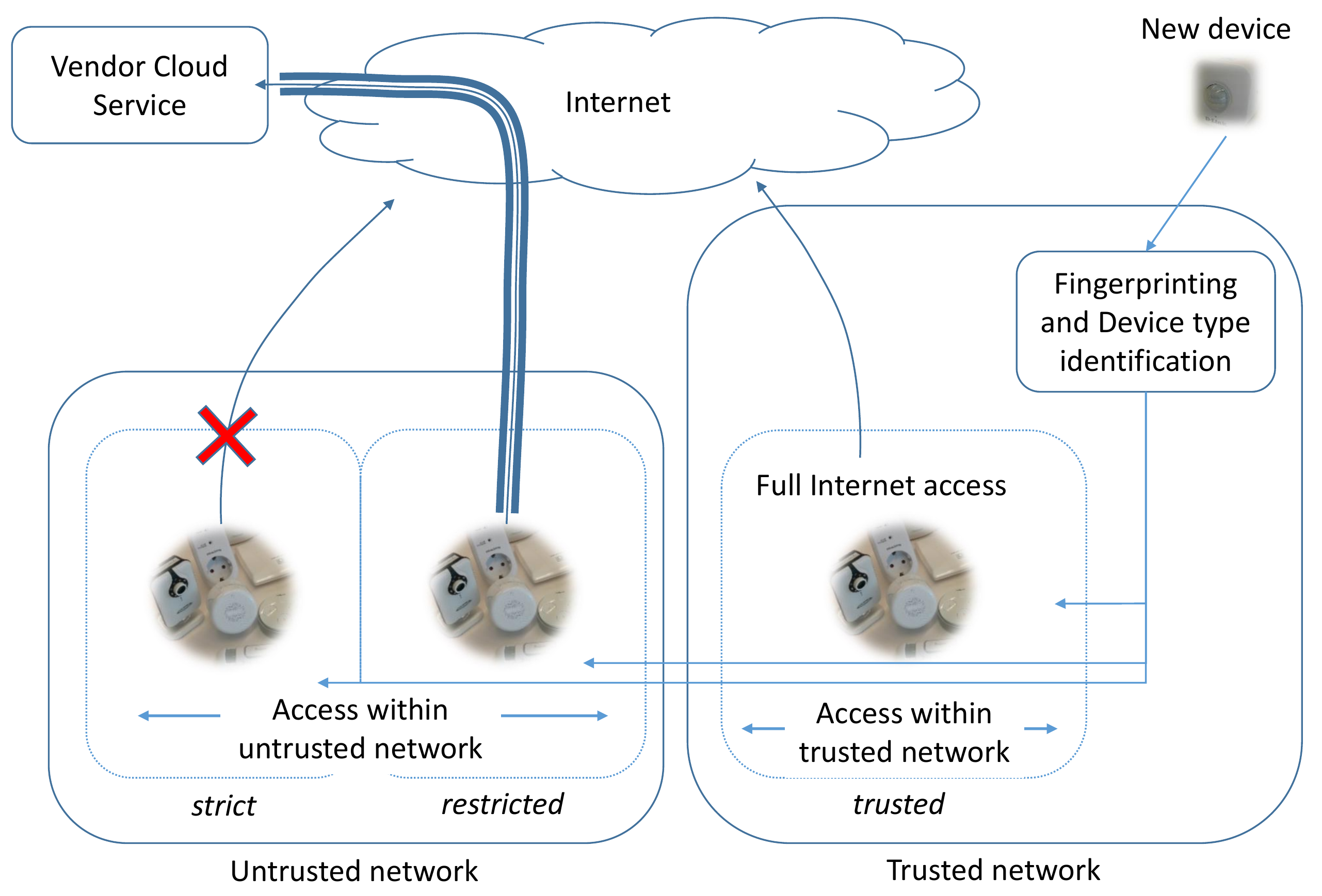}
	\caption{After fingerprinting and device-type identification, new devices are assigned to isolation level \textit{strict}, \textit{restricted} or \textit{trusted}.}
	\label{fig:isolation-levels}
\end{figure}

\begin{itemize}
	\item \textbf{\textit{Strict}} isolation level only allows the device to communicate with other devices in the untrusted network overlay with no Internet access for the device.
	\item \textbf{\textit{Restricted}} isolation level allows the device to communicate with other devices in the untrusted network overlay as well as with a limited set of remote destinations on the Internet (e.g., the vendor's cloud service).
	\item \textbf{\textit{Trusted}} isolation level allows the device to communicate with any other device in the trusted network overlay and unrestricted Internet access.
\end{itemize}

For the \textit{Restricted} isolation level, the \securityservice provides to \securebox an additional set of IP addresses (or DNS names) with which communications are allowed. \securebox stores the received information in local cache. This information can be updated by regular update queries to the \securityservice.
\securebox uses this information to generate enforcement rules to enforce device-specific isolation level required in the network. We identify traffic to/from any device using device MAC addresses, assuming that IoT devices use static MAC addresses. Fig.~\ref{fig:enforcement-rule} shows a sample enforcement rule generated for a \textsl{Device X} identified with MAC address \textsl{13-73-74-7E-A9-C2}. The isolation level for this device is \textit{Restricted}. Therefore, the enforcement rule contains a set of permitted IP addresses through which the device can access its cloud services. These rules are enforced in the network using Open vSwitch (OVS)~\cite{openvswitch-nsdi-15} managed by our customized SDN controller. 

Network isolation at device level granularity ensures that no vulnerable device, when compromised, is able to infect other devices in the trusted network. Our custom module in the SDN controller intercepts all traffic flows in the network and ensures that they are filtered according to the required isolation level assigned to the device by \securityservice. 

Our implementation allows us to extend the traffic filtering mechanism in \securebox to make network isolation even more specific, up to the level of individual flows. \securebox may require a large set of enforcement rules to setup device specific security preferences in the network. However, for any given flow, there is only one matching enforcement rule. In order to minimize the latency experienced during traffic filtering (i.e., time required to find matching enforcement rule for a given flow), enforcement rules are stored in a hash table structure to minimize the lookup time as the enforcement rule cache grows. 

Typically, the traffic between two wireless clients connected to same AP is bridged between the clients and does not go to the routing plane.
Due to this limitation, the traffic cannot be controlled using OVS. 
Some managed networks using proprietary solutions, e.g., Cisco hardware and services, allow network administrators to choose the way traffic is managed on these wireless bridges. However, such functionality is not available in typical consumer wireless APs deployed in existing networks. 
 To enable traffic filtering between wireless devices on typical APs, we leverage \textit{Wireless isolation} feature available on modern wireless APs~\cite{wireless-isolation}.
We use OpenWRT and Linux drivers to redirect traffic between wireless clients through OVS for identifying the devices connected to the wireless interface. This allows us to manage device-to-device communications and maintain required isolation between wireless devices.

\section{Evaluation}
\label{sect:evaluation}
\begin{figure}
	\centering
	\includegraphics[width=.8\columnwidth]{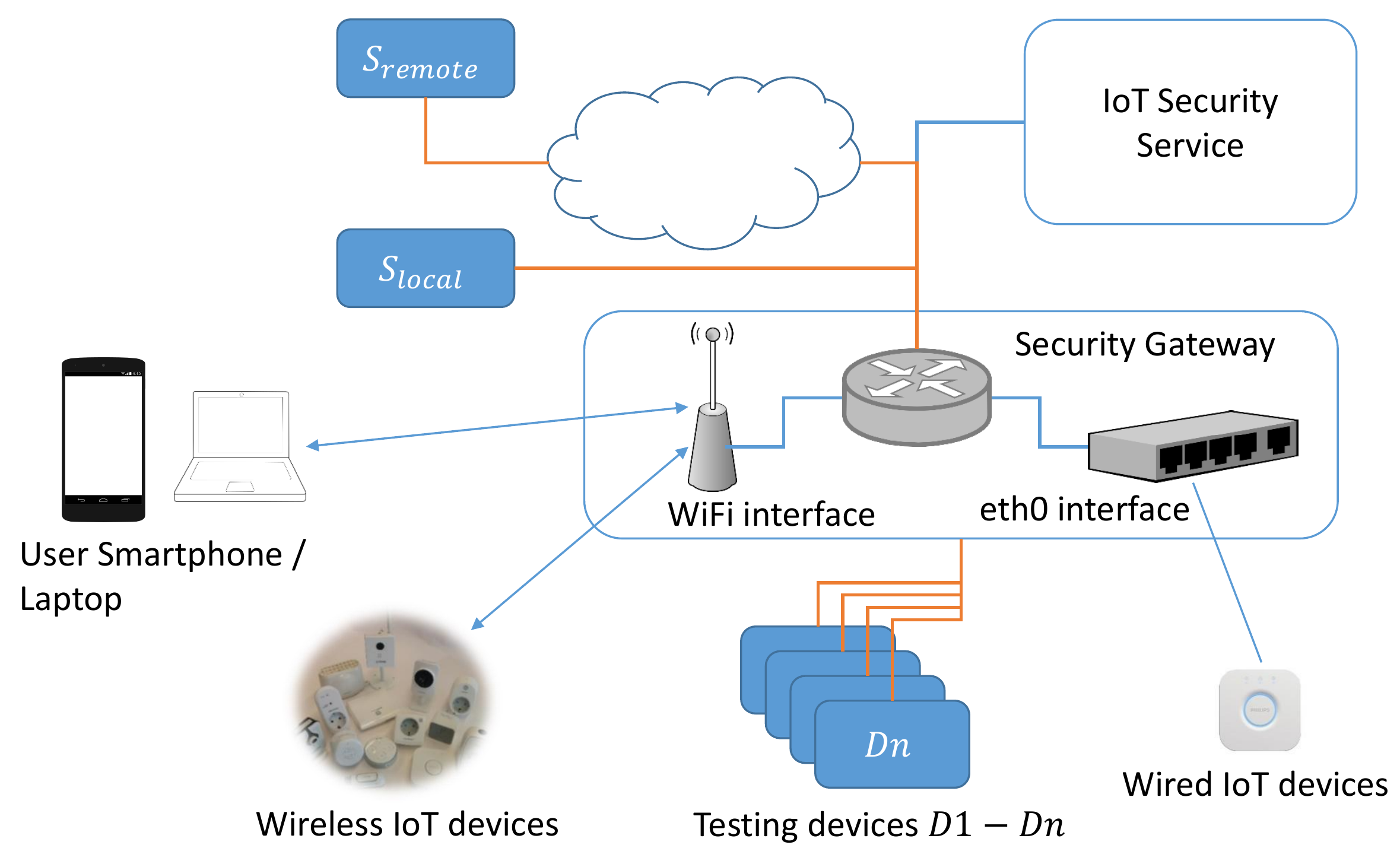}
	\caption{Lab setup for IoT device monitoring and mitigation enforcement.}
	\label{fig:lab-setup}
\end{figure}

To evaluate the IoT device identification and enforcement of mitigation measures, we developed a prototype system in an IoT laboratory environment shown in Fig.~\ref{fig:lab-setup} in order to be able to simulate the IoT device setup process in typical home and small office settings and test the performance of the enforcement of mitigation measures. We collected traffic measurements about the packets sent by each device during the setup process and used the measurements in building the classification model for IoT device identification as described in Sect.~\ref{sect:identification}.

\subsection{Device Fingerprint Collection}

Measurement collection for device fingerprinting was implemented with a Linux Laptop running Kali Linux. The software package \texttt{hostapd} was used to set up a WiFi access point on the laptop emulating the WiFi interface on a WiFi AP. Similarly, an external Ethernet interface was connected to the laptop for emulating the Ethernet ports typically present on APs.
The packet capture module was implemented by using \texttt{tcpdump} on the monitored WiFi and Ethernet interfaces so that all network traffic visible to the \securebox on both wireless and wired network interfaces could be recorded and forwarded to the fingerprinting module.
Data collection was controlled by a scripted UI showing the test person performing the device setup process the necessary step-by-step instructions required to complete the connection of each device to the network. 
The test scripts were manually compiled following the printed or on-line user manual of each device.
The setup process of most of the examined IoT devices was facilitated with a smartphone app or in a few cases a PC application. For such devices, the corresponding app was installed on a testing smartphone or laptop and used in the setup process according to the provided instructions.

\subsubsection{Tested devices}
\label{sect:testdevices}

A representative set of IoT devices targeted for regular consumers that were available in the European market during Q1 2016 was selected for our experiments. These covered most common device classes related to smart lighting, home automation, security cameras, household appliances and health monitoring devices. Most of the tested devices were connected to the user's network via WiFi or Ethernet, but some devices used other IoT protocols like ZigBee or Z-Wave to connect to the network indirectly through an Ethernet or WiFi hub device. For such devices, we focused on monitoring the indirect traffic generated by the hub device acting as a gateway towards the user's network. An overview of the tested devices is shown in Tab.~\ref{tab:evaluated-devices}.

\newcommand{\mcrot}[4]{\multicolumn{#1}{#2}{\rlap{\rotatebox{#3}{#4}~}}} 
\newcommand{\bul}{\textbullet}
\newcommand{\cir}{$\circ$}

\begin{table*}
	\centering
	\caption{List of IoT devices used in the evaluation and their supported connectivity technologies}
	\label{tab:evaluated-devices}
	\begin{tabular}[!th]{lllccccc}
		Identifier & Device Model & \mcrot{1}{l}{60}{WiFi} & \mcrot{1}{l}{60}{ZigBee} & \mcrot{1}{l}{60}{Ethernet} & \mcrot{1}{l}{60}{Z-Wave} & \mcrot{1}{l}{60}{Other}\\
		\hline
		Aria & Fitbit Aria WiFi-enabled scale & \bul & \cir & \cir & \cir & \cir \\
		HomeMaticPlug & Homematic pluggable switch HMIP-PS & \cir & \cir & \cir & \cir & \bul \\
		Withings & Withings Wireless Scale WS-30 & \bul & \cir & \cir & \cir & \cir \\
		MAXGateway & MAX! Cube LAN Gateway for MAX! Home automation sensors & \cir & \cir & \bul & \cir & \bul \\
		HueBridge & Philips Hue Bridge model 3241312018  & \cir & \bul & \bul & \cir & \cir \\
		HueSwitch & Philips Hue Light Switch PTM 215Z  & \cir & \bul & \cir & \cir & \cir \\
		EdnetGateway & Ednet.living Starter kit power Gateway  & \bul & \cir & \cir & \cir & \bul \\
		EdnetCam & Ednet Wireless indoor IP camera Cube  & \bul & \cir & \bul & \cir & \cir \\
		EdimaxCam & Edimax IC-3115W Smart HD WiFi Network Camera  & \bul & \cir & \bul & \cir & \cir \\
		Lightify & Osram Lightify Gateway & \bul & \bul & \cir & \cir & \cir \\
		WeMoInsightSwitch & WeMo Insight Switch model F7C029de & \bul & \cir & \cir & \cir & \cir \\
		WeMoLink & WeMo Link Lighting Bridge model F7C031vf & \bul & \bul & \cir & \cir & \cir \\
		WeMoSwitch & WeMo Switch model F7C027de& \bul & \cir & \cir & \cir & \cir \\
		D-LinkHomeHub & D-Link Connected Home Hub DCH-G020 & \bul & \cir & \bul & \bul & \cir\\
		D-LinkDoorSensor & D-Link Door \& Window sensor & \cir & \cir & \cir & \bul & \cir\\
		D-LinkDayCam & D-Link WiFi Day Camera DCS-930L & \bul & \cir & \bul & \cir & \cir\\
		D-LinkCam & D-Link HD IP Camera DCH-935L & \bul & \cir & \cir & \cir & \cir \\
		D-LinkSwitch & D-Link Smart plug DSP-W215 & \bul & \cir & \cir & \cir & \cir \\
		D-LinkWaterSensor & D-Link Water sensor DCH-S160 & \bul & \cir & \cir & \cir & \cir \\
		D-LinkSiren	& D-Link Siren DCH-S220 & \bul & \cir & \cir & \cir & \cir \\
		D-LinkSensor & D-Link WiFi Motion sensor DCH-S150 & \bul & \cir & \cir & \cir & \cir \\
		TP-LinkPlugHS110 & TP-Link WiFi Smart plug HS110 & \bul & \cir & \cir & \cir & \cir \\
		TP-LinkPlugHS100 & TP-Link WiFi Smart plug HS100 & \bul & \cir & \cir & \cir & \cir \\
		EdimaxPlug1101W & Edimax SP-1101W Smart Plug Switch & \bul & \cir & \cir & \cir & \cir \\
		EdimaxPlug2101W & Edimax SP-2101W Smart Plug Switch & \bul & \cir & \cir & \cir & \cir\\
		SmarterCoffee & Smarter SmarterCoffee coffee machine SMC10-EU & \bul & \cir & \cir & \cir & \cir \\
		iKettle2 & Smarter iKettle 2.0 water kettle SMK20-EU & \bul & \cir & \cir & \cir & \cir\\
	\end{tabular}
\end{table*}

%

For each tested device, the typical device setup process was repeated $n=20$ times in order to generate sufficient fingerprints for classification model training. After each testing round, a hard reset of the tested device was performed to return it to its default factory settings.
Typically, a setup procedure for a device involved activating the device, connecting to the device directly over WiFi (the device sets up an ad-hoc WiFi access point) or Ethernet with the help of a vendor-provided app and transmitting WiFi credentials to the user's network over this connection to the device. After this, the device would typically reset and connect to the user's network using the provided credentials. During this setup procedure, all network traffic visible to the \securebox was recorded and provided to the fingerprinting module for further processing. The fingerprints generated were then transferred to the \securityservice for off-line training of the classification model. The dataset collected from our evaluation setup is available on request for research use.

\begin{figure}
	\includegraphics[width=\columnwidth]{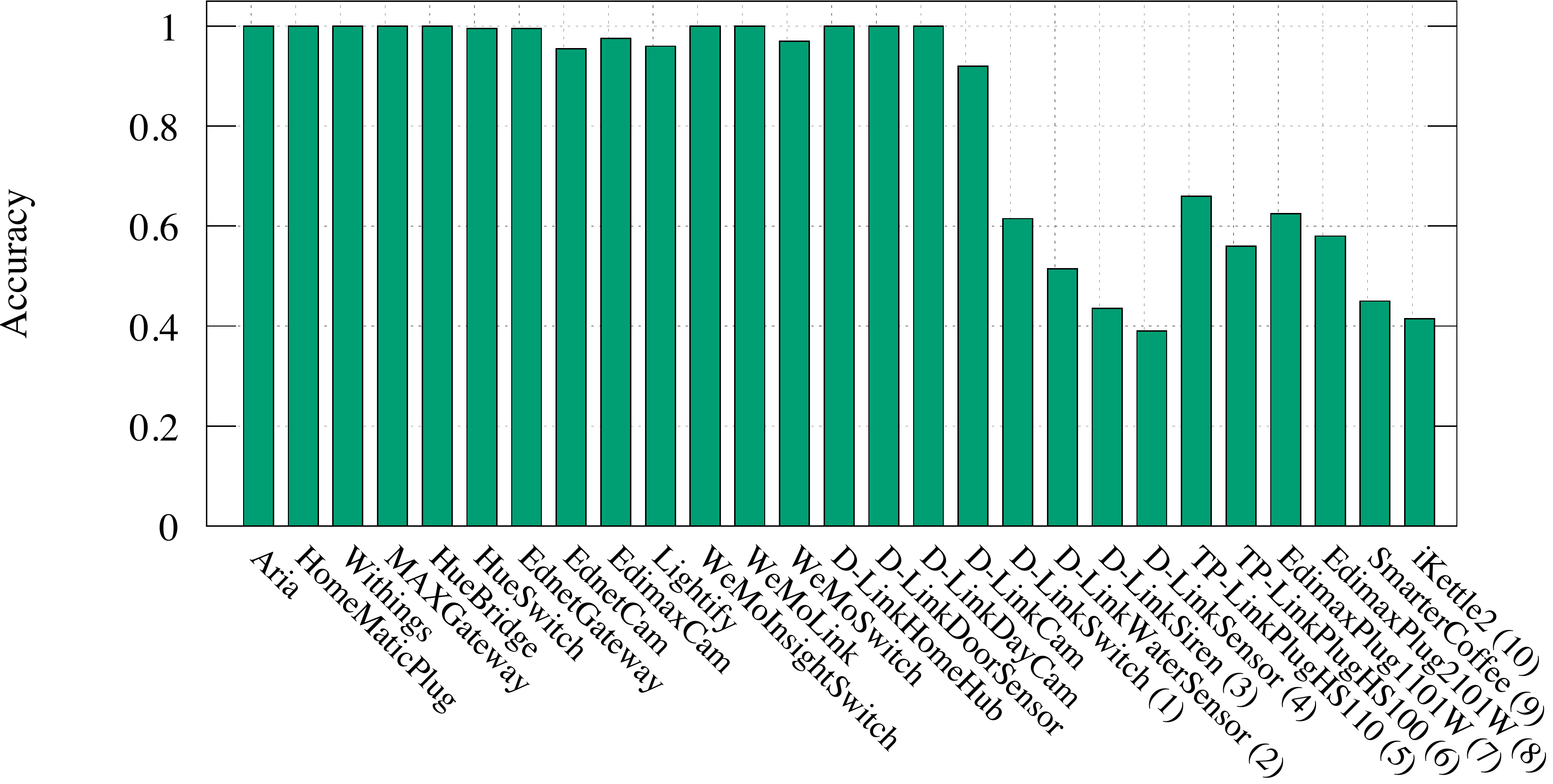}
	\caption{Ratio of correct identification for 27 device-types}
	\label{fig:accuracy}
\end{figure}

\subsection{IoT Device Identification}
\label{sect:identification_eval}

The fingerprints $\mathbf{F}$ and $\mathbf{F'}$ were extracted from the traffic captures following the technique introduced in Sect.~\ref{sect:fingerprinting} to obtain a dataset of 540 fingerprints representing 27 device-types.
The IoT device identification method was evaluated through a stratified 10-fold cross-validation process using this dataset. 
At each fold, we used the training data to learn one classification model per device-type taking all the \textit{n} fingerprints $\mathbf{F'}$ of the targeted type as one class and \textit{10*n} randomly selected fingerprints $\mathbf{F'}$ from the rest to represent the other class.
The testing data were subjected to the 27 learned models to get a prediction from each. In case of positive decision from several classifiers, edit distance discrimination was performed using fingerprints $\mathbf{F}$ randomly selected from the training data as described in Section~\ref{sect:classification}.
The cross-validation was repeated 10 times to generalize the results.

The ratio of correct identification for each device-type is reported in Fig.~\ref{fig:accuracy}. The accuracy of identification is over $0.95$ for 17 devices, most of them reaching $1$. However, we can see that 10 devices are correctly identified with an accuracy around $0.5$, which is lower but still good considering a random type assignment that would give $1/27= 0.037$ accuracy. The global ratio of correct identification over the 27 devices is $0.815$.
To better understand why the 10 aforementioned devices got identified with lower accuracy than others, Table~\ref{tab:confusion} depicts their confusion matrix. We can see that the misidentification occurs between similar devices from same vendors, i.e., a set of similar D-Link devices (1-4), two models of smart plugs from TP-Link (5-6) and Edimax (7-8) and a coffee machine (9) and water kettle (10) from the same vendor.
In contrast, Fig.~\ref{fig:accuracy} shows that our identification technique is able to distinguish devices from same vendor with different purposes, e.g., D-Link camera, hub and sensors, WeMo devices, Edimax camera and plug, etc.
From a system point of view, the misidentified devices are very similar: D-Link water sensor (2), siren (3) and sensor (4) have identical hardware and firmware version, as TP-Link plugs (5-6) do. Hence, these devices are likely to share vulnerabilities, if any. 
Moreover, SmarterCoffee (9) and iKettle2 (10) received a firmware update during the data collection period that led to generate new fingerprints. These fingerprints were distinguishable from the one generated with their older firmware version. This suggests that our device fingerprint is able to discriminate device-types as long as they have different hardware/firmware versions, and vulnerability patching would change the fingerprint of a device.
Hence, this specific misidentification issue is not a concern for our purpose of identifying vulnerable devices.  

\begin{table}[tbh]
\caption{Confusion matrix for 10 devices with low identification rate (device index find corresponding names in Fig.~\ref{fig:accuracy}) \newline  A= actual type / P= predicted type} \label{tab:confusion}

\centering
\begin{tabular}{c | c c c c c c c c c c }

\textbf{A\textbackslash P} & \textbf{1}  & \textbf{2} & \textbf{3}  & \textbf{4} & \textbf{5}  & \textbf{6} & \textbf{7}  & \textbf{8} & \textbf{9}  & \textbf{10} \\ \hline
\textbf{1} & 123 & 23 & 28 & 26 & 0 & 0 & 0 & 0 & 0 & 0 \\
\textbf{2} & 0 & 103 & 42 & 55 & 0 & 0 & 0 & 0 & 0 & 0  \\
\textbf{3} & 4 & 55 & 87 & 54 & 0 & 0 & 0 & 0 & 0 & 0 \\
\textbf{4} & 8 & 65 & 49 & 78 & 0 & 0 & 0 & 0 & 0 & 0  \\
\textbf{5} & 0 & 0 & 0 & 0 & 132 & 68 & 0 & 0 & 0 & 0  \\
\textbf{6} & 0 & 0 & 0 & 0 & 88 & 112 & 0 & 0 & 0 & 0  \\
\textbf{7} & 0 & 0 & 0 & 0 & 0 & 0 & 125 & 75 & 0 & 0  \\
\textbf{8} & 0 & 0 & 0 & 0 & 0 & 0 & 84 & 116 & 0 & 0  \\
\textbf{9} & 0 & 0 & 0 & 0 & 0 & 0 & 0 & 0 & 90 & 110  \\
\textbf{10} & 0 & 0 & 0 & 0 & 0 & 0 & 0 & 0 & 117 & 83  \\ 
\end{tabular}
\end{table}

From a performance perspective, Tab.~\ref{tab:time} reports the time taken for device-type identification. We see that most of the time is spent on cases where discrimination using edit distance was required. During experiments, 55\% of the analyzed fingerprints matched more than one type and needed a discrimination step that involved between two and five types. On average, seven edit distance computations were needed per device. The average time for device-type identification is around 150 ms. For comparison, the time taken for device setup was between one and two minutes, the packet collection was performed in parallel of this operation. The classification with Random Forest takes very little time ($<$1 ms) and grows linearly with the number of types to identify. This shows that \ourname can easily scale to thousands of device-types while keeping classification time below 100 ms and type identification likely below 1 second.

\begin{table}[tbh]
\caption{Time consumption for device-type identification. Time for single steps is presented at the top and time for an average type identification in our lab setup is presented below.} \label{tab:time}

\centering
\begin{tabular}{l | r r  }

\textbf{Steps} & \textbf{Mean ($\pm$StDev)}    \\ \hline
1 Classification (Random Forest) &  0.014 ms  ($\pm 0.003$)  \\
1 Discrimination (edit distance) & 23.36 ms ($\pm 24.37$)  \\ \hline
Fingerprint extraction & 0.850 ms ($\pm 0.698$)  \\
27 Classifications (Random Forest) & 0.385 ms ($\pm 0.081$)  \\
7 Discriminations (edit distance)  &  156.5 ms ($\pm  170.6$) \\ \hline
\textbf{Type Identification} & \textbf{157.7 ms ($\pm 171.4$)}  \\ 

\end{tabular}
\end{table}

\subsection{Mitigation Measures Enforcement}
Fig.~\ref{fig:lab-setup} shows the lab setup used for testing enforcement mechanism employed by \securebox. In general, \securebox can be deployed in two setups.
\begin{itemize}
\item A Raspberry PI 2 (R-Pi II) device running both the OVS and custom SDN controller acts as standalone \securebox in the network. An external USB WiFi dongle, or integrated WiFi and \texttt{hostapd} used to emulate wireless interface on R-Pi II~\cite{rpi-wifi-ap}.
\item  An off-the-shelf router on which we setup OpenFlow enabled Access Point (OF-AP) by installing OVS software package over the stock \texttt{OpenWRT} OS of the router ~\cite{wireless-isolation}. In that case, the OVS on OF-AP will be managed by our custom SDN controller running on a separate machine.
\end{itemize}

\begin{figure*}[th]
\subfloat[\textbf{Latency increase relative to concurrent flows.}]{\includegraphics[width=0.32\textwidth]{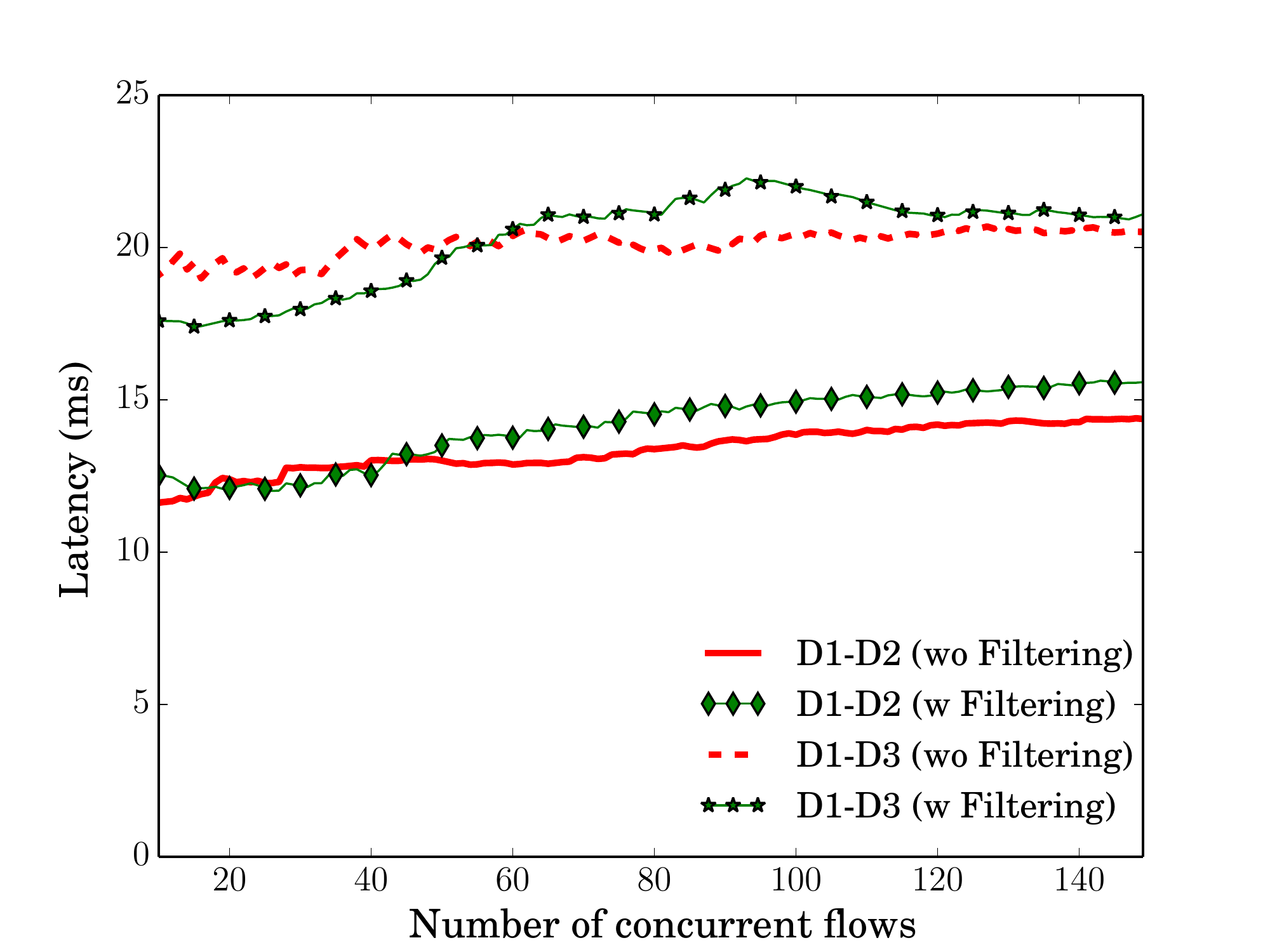}\label{fig:lat-v-flows}}
\hspace{0.1cm}
\subfloat[\textbf{CPU utilization for \securebox.}]{\includegraphics[width=0.32\textwidth]{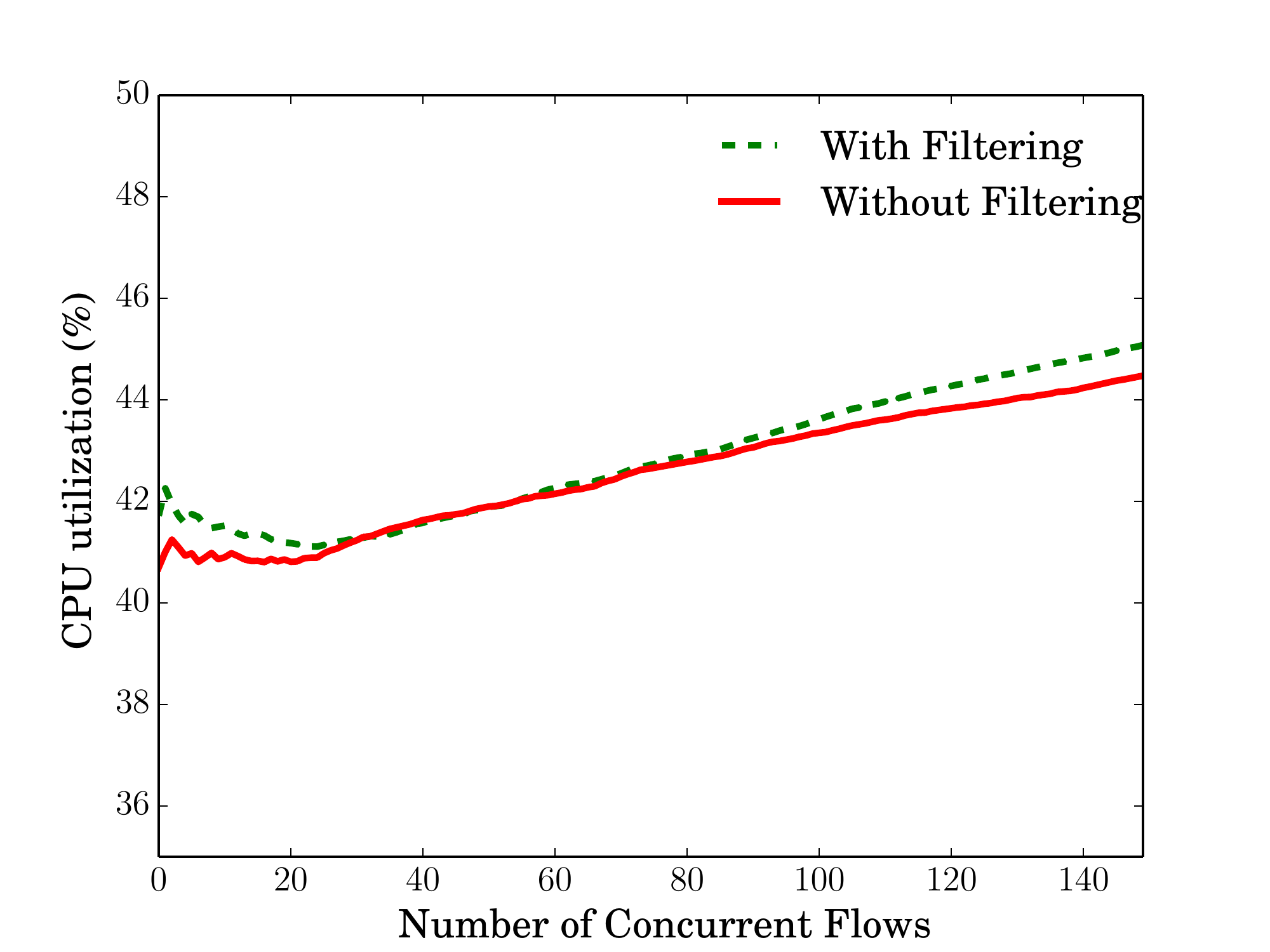}\label{fig:cpu_util}}
\hspace{0.1cm}
\subfloat[\textbf{Memory utilization of \securebox.}]{\includegraphics[width=0.32\textwidth]{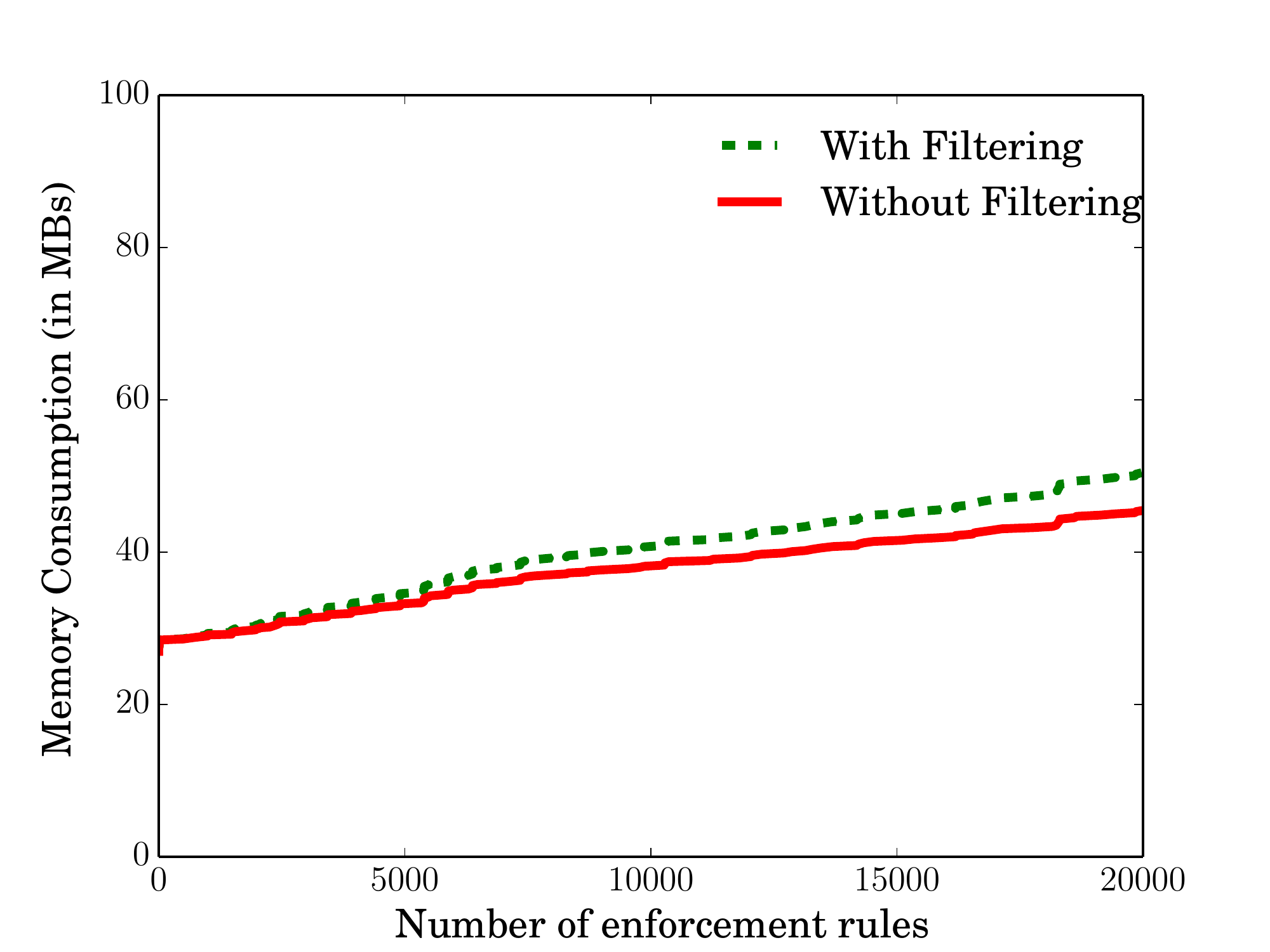}\label{fig:mem_util}}

\caption{\textbf{\ourname Performance evaluation for Raspberry PI based deployment of \securebox.} \textsl{There is only minimal increase in CPU and memory utilization of \securebox with filtering mechanism enabled. Additionally, the increase in latency experienced relative to increasing number of concurrent flows in the network is insignificant in terms of user experience.}}
\label{fig:eval}
\end{figure*}

We selected the first setup for evaluation. For each experiment, we performed 15 iterations for each measured device pair. We measured the latency experienced between devices $D_{1} - D_{n}$ connected to \securebox wireless interface, as well as between devices and servers, where $S_{local}$ is in the local network and $S_{remote}$ is a remote server deployed in Amazon EC2. Table~\ref{tbl:latency_enforcement} shows that the enforcement mechanism employed for traffic filtering by \securebox does not impact the latency experienced by the user. 
Fig.~\ref{fig:lat-v-flows} shows the impact on latency experienced by devices regarding the total number of concurrent flows in the network. The results show that the increase in latency for up to 150 concurrent flows is insignificant to affect user experience or device operations.

\begin{table}[]
\centering
\caption{\textbf{Latency (ms)} experienced by users where $D1-D4$ are user devices connected to \securebox (SGW) and $S_{local}$ and $S_{remote}$ are servers.}
\label{tbl:latency_enforcement}
\begin{tabular}{p{0.7cm}p{1.5cm}p{2.2cm}p{2.2cm}}
\textbf{Source} 	& \textbf{Destination} 	& \textbf{Filtering} \newline \textbf{Mean ($\pm$ StDev)} & \textbf{No Filtering} \newline \textbf{Mean ($\pm$ StDev)} \\
\hline
\textbf{$D1$}     	& $D4$        				& 24.8 ($\pm 1.4$)                     & 24.5 ($\pm 1.4$)                  \\
\textbf{}       	& $S_{local}$        	& 18.4 ($\pm 1.3$)               & 18.2 ($\pm 1.3$)                  \\
\textbf{}			& $S_{remote}$        	& 20.6 ($\pm 3.3$)                     & 20.3 ($\pm 3.1$)                    \\
\textbf{$D2$}     	& $D4$       			    & 28.5 ($\pm 1.7$)                       & 28.2  ($\pm 1.6$)                    \\
\textbf{}       	& $S_{local}$       		& 17.2 ($\pm 1.2$)                     & 17.0 ($\pm 1.2$)                     \\
\textbf{}      	 	& $S_{remote}$        	& 20.0 ($\pm 2.9$)                      & 19.8 ($\pm 3.1$)                    \\
\textbf{$D3$}   	& $D4$        				& 27.6 ($\pm 1.6$)                      & 27.5 ($\pm 1.6$)                   \\
\textbf{}	    	& $S_{local}$        	& 15.5 ($\pm 1.2$)                       & 15.4 ($\pm 1.1$)                    \\
\textbf{}       	& $S_{remote}$        	& 20.6 ($\pm 3.2$)                      & 19.9 ($\pm 3.2$)                   \\
\end{tabular}
\end{table}

\begin{wraptable}{r}{0.27\textwidth}
\caption{Overhead due to filtering mechanism.}\label{tbl:stats-fig6} \centering
\footnotesize
\begin{tabular}{ll}
\multirow{2}{*}{\textbf{Case} }				& \textbf{Overhead}   \\
 & \textbf{Mean ($\pm$ StDev)} \\
\hline
$D1D2$ Latency     	& +5.84\% ($\pm 4.76\%$)    	\\
$D1D3$ Latency   	& +0.71\% ($\pm 5.88\%$)    	\\
{CPU utilization}	& +0.63\% ($\pm 1.8\%$)  		\\
Memory usage	& +7.6\% ($\pm 4.6\%$)   	\\
\end{tabular}
\vspace*{-5pt}
\end{wraptable}

We also measured the memory and CPU overhead of our enforcement mechanism. Fig.~\ref{fig:cpu_util} shows that there is very little overhead in terms of CPU utilization due to traffic filtering mechanism. The overall CPU utilization measurements also shows that \securebox does not require high processing power to perform network operations. 

Similarly, Fig.~\ref{fig:mem_util} shows the amount of memory utilized by \securebox with and without using filtering mechanism is almost similar. The amount of memory used for storing enforcement rules can be controlled by limiting the size of enforcement rule cache and removing unused enforcement rules (for the devices that are no longer connected to the network) from the cache. The percentage increase in memory, CPU utilization as well as latency experienced by user  due to filtering mechanism is given in Table~\ref{tbl:stats-fig6}.  These results show that the overhead on memory and CPU utilization is low. 
Fig.~\ref{fig:eval} also shows that a small form factor PC such as Raspberry Pi II provides sufficient computational resources for a typical environment hosting, e.g., a hundred IoT devices generating the same number of concurrent flows and requiring as many enforcement rules.


\section{Related Work}
\label{sect:relatedwork}
\subsection{Securing IoT Device-to-Device Communications}
\label{sect:soa_d2dcomm}

Authentication schemes tailored for resource constrained devices \cite{7012039} are primarily used to control communications between IoT devices.
Zenger et al. \cite{Zenger2016} proposed a vicinity based pairing mechanism that delegates trust from one node to another based on physical proximity. Messaging between devices can be authenticated using multiple communication channels (e.g. Bluetooth + NFC) to ensure secure pairing \cite{6767206}. 
However, these schemes require some implementation on all devices of the system to be applicable, failing to cope with the IoT brownfield of legacy devices already deployed. 

At run time, communications between IoT devices can be restricted based on high level user requirements that a system, namely SIFT \cite{Liang:2015:SBI}, translates to low level access control policies.
Fully automated techniques to identify malicious communications rely on intrusion detection systems tailored for IoT scenarios \cite{Raza20132661}.
Verification of data sent from a given device can be based on measurement correlation with other devices to identify malicious nodes attempting to pollute measurements \cite{Gisdakis:2015:SDV}. Action can then be taken to discard communications from this device.
The main difference between \ourname and these techniques is that the former is preventive, mitigating the threat of vulnerable devices  when they are inducted in the system and before any malicious communication is initiated. 

A recently announced commercial product that follows a similar conceptual approach as \ourname is F-Secure SENSE~\cite{F-Secure:2016:sense}. However, SENSE is focused on traditional protection technologies like anti-virus capabilities and blocking of botnet controllers and malicious websites~\cite{F-Secure:2016:sense-features}. Unlike \ourname, it does not seem to provide means for automatic identification of arbitrary IoT device-types. 

\subsection{Device Fingerprinting}
\label{sect:soa_fingerprinting}

Early work in 802.11 wireless communication fingerprinting targeted the identification of hardware and driver specific characteristics, with active \cite{Bratus:2008:active} or passive methods \cite{cache2006fingerprinting,Franklin:2006:passive,maurice2013improving}. Cache \cite{cache2006fingerprinting} used 802.11 frames' duration field that only takes few discrete values depending on driver implementation to identify WiFi drivers. 
Passively recording 802.11 probing frames inter-arrival time from a device, Franklin \textit{et al.} \cite{Franklin:2006:passive} were able to classify 17 WiFi drivers with an accuracy ranging between 77\% and 96\% using a Bayesian classification method.
While relying on passively captured network traffic as we do, these techniques build hardware/driver specific fingerprints that are too coarse-grained for our purpose of identifying device-types. Low cost IoT devices are likely to use identical cheap WiFi interfaces and corresponding drivers leading to aggregate a wide range of device-types in the same class using these techniques.

On the other hand, network features such as packet destination, SSID probes, broadcast packet size and MAC protocol fields were used to fingerprint specific users in \cite{Pang:2007:user}. 
Hardware specific characteristics such as clock skew \cite{Kohno2005,Jana:2008:fast,Arackaparambil:2010:reliability} or radio-frequency signature \cite{Ureten2007,Brik:2008:wireless} can be used to identify a unique network interface card, mostly for rogue wireless Access Point detection purposes.
IoT-specific techniques target mostly high-end devices, leveraging mobile device configuration \cite{kurtz2016fingerpritning}, or sensor specific features \cite{bojinov2014mobile,Bertini:2015:Profile,VanGoethem2016} though similar techniques can be applied to larger classes of IoT devices \cite{Sharaf2016on,Haider:2016:Trusted}. However, sensor data analysis only addresses the identification of a limited class of devices actually reporting such information.
All these methods build fingerprints able to uniquely identify a device, which is too specific to identify an unknown device as belonging to one type. Our technique is positioned between the former and latter approaches, providing the right granularity to identify device-types from passive traffic captures. 

Gao \textit{et al.} \cite{Gao:2010:passive} similarly introduced a passive technique to identify device-types using the fact that a type of device modifies a packet in a unique way, due to its internal architecture, while processing it. Capturing incoming and outgoing packets and applying wavelet analysis, they were able to discriminate device-types. This technique only applies to devices processing and forwarding packets such as routing devices but is not applicable to end point IoT devices that we target in our work.


GTID \cite{Radhakrishnan2015} addresses as well device-type identification. GTID builds a feature vector composed of inter-arrival time of packets sent by a device for a specific type of traffic (e.g. Skype, ICMP, etc.). Feature vectors are used in a neural network predicting as many classes as there are device-types to identify. The main difference with our work is that our fingerprints are not specific to a type of traffic sent at high rate over a significant period of time. In contrast to devices used for GTID evaluation, i.e., smartphones and tablets, most IoT devices generate less traffic with less diversity limiting this approach to high-end devices. Our technique has a wider scope, applying to wireless and wired traffic. In addition, using a single multi-class neural network model in GTID requires full model relearning when one new type is identified while our ``one classifier per type approach'' does not.

\section{Discussion and Future Work}
\label{sect:discussion}
\subsection{Support for Legacy Installations}
\label{sect:legacy}

In this work we considered a setting in which the user introduces new devices to his or her network via the \securebox and device fingerprinting and identification is based on the characteristic traffic patterns of devices when they are initially connected to the network. 
However, we envisage \ourname also to support legacy network installations by updating a compatible gateway router with \securebox functionality with the necessary fingerprinting and traffic filtering components by means of a software update.

In such cases, fingerprinting and device-type identification happens after devices have already been connected to the user's network. The profiling of devices has therefore to happen based on the communication behaviour that devices exhibit during standby (e.g., heartbeat messages to the vendor's cloud solution), or, during the normal operation of the device. Our working hypothesis is that similarly as with the initial setup messages, message exchanges during standby and operation cycles are likely to be characteristic for particular device-types and therefore form a good basis for device-type identification. In future work, we intend to investigate this approach further to verify its effectiveness.

Legacy network installations often use WPA2-Personal with a network-specific PSK to authenticate devices to the network. If there are devices with security vulnerabilities present, it is possible that the confidentiality of the PSK is breached, leaking the PSK to an adversary. In order to protect the user's IoT devices against a local adversary using a leaked PSK to infiltrate the user's network, we need to establish a virtual network overlay consisting of an untrusted network representing the existing potentially vulnerable network setup and a new, trusted network. All devices in the legacy installation will be initially in the untrusted network. After performing device-type identification, such devices that do not have known vulnerabilities will be transferred to the trusted network by using the re-keying functionality defined by WiFi Protected Setup (WPS)~\cite{WiFiAlliance2014}. For devices supporting WPS re-keying, it can be triggered by deprecating the legacy network's WPA2-Personal PSK, after which the device will trigger a re-keying exchange to obtain a new WPA2-Personal PSK for the trusted network overlay. The new PSK provided during re-keying will be device-specific, in order to allow device-specific authentication in the trusted network overlay.

For legacy devices that do not support WPS re-keying, there are two options: 1) \emph{WPA2-Personal PSK remains in force}: The device continues to operate in the untrusted network, but is unable to communicate with devices in the trusted network. 2) \emph{WPA2-Personal PSK is deprecated}: The device will be unable to access the user's network. Manual re-introduction of the device to the user's network is required before it can re-join. The advantage of the former solution is that it does not impact user experience by disconnecting devices from the user's network. However, devices remaining in the untrusted network are likely more exposed to adversarial attacks. Therefore, the \securebox may optionally prompt the user (e.g., through its management interface) to re-authenticate such devices manually. Devices with known security vulnerabilities will remain in the untrusted network. The trusted and untrusted networks, however, will be strictly separated from each other.

\subsection{Impact of Software Updates}
In this work, we defined a device-type to denote the combination of a device's make, model and software version. As in our set of test devices, however, only a few devices offered the possibility for a software update during our experimentation period, we were not able to comprehensively investigate this capability. For three devices for which updates were applied, these updates led to generate distinguishable fingerprints between software versions of these devices. In our future work we expect to be able to investigate this further, as we expect over time software updates to become available for a larger set of the analysed devices.

\section{Conclusion}
\label{sect:conclusion}
In this paper, we presented \ourname, a system for automatic identification and security enforcement of IoT devices with potential security vulnerabilities. We propose a novel approach in which identification is based on profiling of the device type-specific communication behaviour of individual devices. Our system provides protection for the user's network by enforcing network isolation where communications of potentially vulnerable devices are strictly controlled, thereby effectively mitigating security risks related to these devices.

\bibliographystyle{IEEEtran}
\bibliography{iotlab,IEEEabrv}

\begin{thebibliography}{10}
\providecommand{\url}[1]{#1}
\csname url@samestyle\endcsname
\providecommand{\newblock}{\relax}
\providecommand{\bibinfo}[2]{#2}
\providecommand{\BIBentrySTDinterwordspacing}{\spaceskip=0pt\relax}
\providecommand{\BIBentryALTinterwordstretchfactor}{4}
\providecommand{\BIBentryALTinterwordspacing}{\spaceskip=\fontdimen2\font plus
\BIBentryALTinterwordstretchfactor\fontdimen3\font minus
  \fontdimen4\font\relax}
\providecommand{\BIBforeignlanguage}[2]{{%
\expandafter\ifx\csname l@#1\endcsname\relax
\typeout{** WARNING: IEEEtran.bst: No hyphenation pattern has been}%
\typeout{** loaded for the language `#1'. Using the pattern for}%
\typeout{** the default language instead.}%
\else
\language=\csname l@#1\endcsname
\fi
#2}}
\providecommand{\BIBdecl}{\relax}
\BIBdecl

\bibitem{Businessinsider2016}
\BIBentryALTinterwordspacing
J.~Greenough. How the {'Internet of Things}' will impact consumers, businesses,
  and governments in 2016 and beyond. Business Insider. [Accessed: 2016-07-18].
  [Online]. Available:
  \url{http://www.businessinsider.de/how-the-internet-of-things-market-will-grow-2014-10}
\BIBentrySTDinterwordspacing

\bibitem{Core2013}
{Core Security}, ``{AVTECH DVR} multiple vulnerabilities,''
  \url{http://www.coresecurity.com/advisories/avtech-dvr-multiple-vulnerabilities},
  Aug. 2013, [Accessed: 2016-03-29].

\bibitem{Pentestpartners2015}
D.~Lodge, ``Hacking a {Wi-Fi} coffee machine -- part 1,''
  \url{https://www.pentestpartners.com/blog/hacking-a-wi-fi-coffee-machine-part-1/},
  2015, [Accessed: 2016-03-29].

\bibitem{Senrio2016}
\BIBentryALTinterwordspacing
Senrio. 400,000 publicly available {IoT} devices vulnerable to single flaw.
  [Accessed: 2016-07-07]. [Online]. Available:
  \url{http://blog.senr.io/blog/400000-publicly-available-iot-devices-vulnerable-to-single-flaw}
\BIBentrySTDinterwordspacing

\bibitem{SECConsult2016}
\BIBentryALTinterwordspacing
{SEC Consult}. (2016, Sep.) House of keys: 9 months later\ldots 40\% worse.
  [Accessed: 2016-09-06]. [Online]. Available:
  \url{http://blog.sec-consult.com/2016/09/house-of-keys-9-months-later-40-worse.html}
\BIBentrySTDinterwordspacing

\bibitem{Sivaraman2016}
V.~Sivaraman, D.~Chan, D.~Earl, and R.~Boreli, ``Smart-phones attacking
  smart-homes,'' in \emph{Proceedings of the 9th ACM Conference on Security and
  Privacy in Wireless and Mobile Networks}, ser. WiSec '16.\hskip 1em plus
  0.5em minus 0.4em\relax ACM, 2016, pp. 195--200.

\bibitem{CVE2016}
\BIBentryALTinterwordspacing
{MITRE Corporation}. Common vulnerabilities and exposures. [Online]. Available:
  \url{https://cve.mitre.org/data/downloads/index.html}
\BIBentrySTDinterwordspacing

\bibitem{Feng:2016:scalable}
Q.~Feng, R.~Zhou, C.~Xu, Y.~Cheng, B.~Testa, and H.~Yin, ``Scalable graph-based
  bug search for firmware images,'' in \emph{Proceedings of the 2016 ACM SIGSAC
  Conference on Computer and Communications Security}, ser. CCS '16.\hskip 1em
  plus 0.5em minus 0.4em\relax ACM, 2016, pp. 480--491.

\bibitem{dingledine2004tor}
R.~Dingledine, N.~Mathewson, and P.~Syverson, ``Tor: The second-generation
  onion router,'' DTIC Document, Tech. Rep., 2004.

\bibitem{Bratus:2008:active}
S.~Bratus, C.~Cornelius, D.~Kotz, and D.~Peebles, ``Active behavioral
  fingerprinting of wireless devices,'' in \emph{Proceedings of the First ACM
  Conference on Wireless Network Security}, ser. WiSec '08.\hskip 1em plus
  0.5em minus 0.4em\relax ACM, 2008, pp. 56--61.

\bibitem{cache2006fingerprinting}
J.~Cache, ``Fingerprinting 802.11 implementations via statistical analysis of
  the duration field,'' \emph{Uninformed}, vol.~5, 2006.

\bibitem{Franklin:2006:passive}
J.~Franklin, D.~McCoy, P.~Tabriz, V.~Neagoe, J.~Van~Randwyk, and D.~Sicker,
  ``Passive data link layer 802.11 wireless device driver fingerprinting,'' in
  \emph{Proceedings of the 15th Conference on USENIX Security Symposium -
  Volume 15}, ser. USENIX-SS'06.\hskip 1em plus 0.5em minus 0.4em\relax
  Berkeley, CA, USA: USENIX Association, 2006.

\bibitem{maurice2013improving}
C.~Maurice, S.~Onno, C.~Neumann, O.~Heen, and A.~Francillon, ``Improving 802.11
  fingerprinting of similar devices by cooperative fingerprinting,'' in
  \emph{Proceedings of the 2013 International Conference on Security and
  Cryptography (SECRYPT)}, 2013, pp. 1--8.

\bibitem{Kohno2005}
T.~Kohno, A.~Broido, and K.~C. Claffy, ``Remote physical device
  fingerprinting,'' \emph{IEEE Transactions on Dependable and Secure
  Computing}, vol.~2, no.~2, pp. 93--108, April 2005.

\bibitem{Jana:2008:fast}
S.~Jana and S.~K. Kasera, ``On fast and accurate detection of unauthorized
  wireless access points using clock skews,'' in \emph{Proceedings of the 14th
  ACM International Conference on Mobile Computing and Networking}, ser.
  MobiCom '08.\hskip 1em plus 0.5em minus 0.4em\relax ACM, 2008, pp. 104--115.

\bibitem{Arackaparambil:2010:reliability}
C.~Arackaparambil, S.~Bratus, A.~Shubina, and D.~Kotz, ``On the reliability of
  wireless fingerprinting using clock skews,'' in \emph{Proceedings of the
  Third ACM Conference on Wireless Network Security}, ser. WiSec '10.\hskip 1em
  plus 0.5em minus 0.4em\relax ACM, 2010, pp. 169--174.

\bibitem{Ureten2007}
O.~Ureten and N.~Serinken, ``Wireless security through rf fingerprinting,''
  \emph{Canadian Journal of Electrical and Computer Engineering}, vol.~32,
  no.~1, pp. 27--33, Winter 2007.

\bibitem{Brik:2008:wireless}
V.~Brik, S.~Banerjee, M.~Gruteser, and S.~Oh, ``Wireless device identification
  with radiometric signatures,'' in \emph{Proceedings of the 14th ACM
  International Conference on Mobile Computing and Networking}, ser. MobiCom
  '08.\hskip 1em plus 0.5em minus 0.4em\relax ACM, 2008, pp. 116--127.

\bibitem{Gao:2010:passive}
K.~Gao, C.~Corbett, and R.~Beyah, ``A passive approach to wireless device
  fingerprinting,'' in \emph{Proceedings of the 2010 IEEE/IFIP International
  Conference on Dependable Systems Networks (DSN)}.\hskip 1em plus 0.5em minus
  0.4em\relax IEEE, 2010, pp. 383--392.

\bibitem{Radhakrishnan2015}
S.~V. Radhakrishnan, A.~S. Uluagac, and R.~Beyah, ``{GTID}: A technique for
  physical device and device type fingerprinting,'' \emph{IEEE Transactions on
  Dependable and Secure Computing}, vol.~12, no.~5, pp. 519--532, 2015.

\bibitem{Pang:2007:user}
J.~Pang, B.~Greenstein, R.~Gummadi, S.~Seshan, and D.~Wetherall, ``802.11 user
  fingerprinting,'' in \emph{Proceedings of the 13th Annual ACM International
  Conference on Mobile Computing and Networking}, ser. MobiCom '07.\hskip 1em
  plus 0.5em minus 0.4em\relax ACM, 2007, pp. 99--110.

\bibitem{He:2009:learning}
H.~He and E.~A. Garcia, ``Learning from imbalanced data,'' \emph{IEEE Trans. on
  Knowl. and Data Eng.}, vol.~21, no.~9, pp. 1263--1284, Sep. 2009.

\bibitem{Breiman2001}
L.~Breiman, ``Random forests,'' \emph{Machine Learning}, vol.~45, no.~1, pp.
  5--32, 2001.

\bibitem{Damerau:1964:technique}
F.~J. Damerau, ``A technique for computer detection and correction of spelling
  errors,'' \emph{Commun. ACM}, vol.~7, no.~3, pp. 171--176, Mar. 1964.

\bibitem{floodlight-controller}
{Big Switch Networks}, ``Project floodlight - floodlight {OpenFlow}
  controller,'' \url{http://www.projectfloodlight.org/floodlight/}, Oct. 2016,
  [Accessed: 2016-09-17].

\bibitem{openvswitch-nsdi-15}
B.~Pfaff, J.~Pettit, T.~Koponen, E.~J. Jackson, A.~Zhou, J.~Rajahalme,
  J.~Gross, A.~Wang, J.~Stringer, P.~Shelar, K.~Amidon, and M.~Casado, ``The
  design and implementation of {Open vSwitch},'' in \emph{Proceedings of the
  12th USENIX Conference on Networked Systems Design and Implementation}.\hskip
  1em plus 0.5em minus 0.4em\relax USENIX Association, 2015, pp. 117--130.

\bibitem{wireless-isolation}
S.~H\"{a}t\"{o}nen, P.~Savolainen, A.~Rao, H.~Flinck, and S.~Tarkoma,
  ``Off-the-shelf software-defined {Wi-Fi} networks,'' in \emph{Proceedings of
  the 2016 Conference on ACM SIGCOMM 2016 Conference}, ser. SIGCOMM '16.\hskip
  1em plus 0.5em minus 0.4em\relax ACM, 2016, pp. 609--610.

\bibitem{rpi-wifi-ap}
\BIBentryALTinterwordspacing
{Lady ada}, ``Setting up a rasperry {PI} as a {WiFi} access point,'' Ada Fruit.
  [Online]. Available:
  \url{https://cdn-learn.adafruit.com/downloads/pdf/setting-up-a-raspberry-pi-as-a-wifi-access-point.pdf}
\BIBentrySTDinterwordspacing

\bibitem{7012039}
J.~L. Hernández-Ramos, M.~P. Pawlowski, A.~J. Jara, A.~F. Skarmeta, and
  L.~Ladid, ``Toward a lightweight authentication and authorization framework
  for smart objects,'' \emph{IEEE Journal on Selected Areas in Communications},
  vol.~33, no.~4, pp. 690--702, 2015.

\bibitem{Zenger2016}
C.~T. Zenger, M.~Pietersz, J.~Zimmer, J.-F. Posielek, T.~Lenze, and C.~Paar,
  ``Authenticated key establishment for low-resource devices exploiting
  correlated random channels,'' \emph{Computer Networks}, pp.~--, 2016.

\bibitem{6767206}
V.~Mora-Afonso, P.~Caballero-Gil, and J.~Molina-Gil, ``Strong authentication on
  smart wireless devices,'' in \emph{Second International Conference on Future
  Generation Communication Technologies (FGCT)}, 2013, pp. 137--142.

\bibitem{Liang:2015:SBI}
C.-J.~M. Liang, B.~F. Karlsson, N.~D. Lane, F.~Zhao, J.~Zhang, Z.~Pan, Z.~Li,
  and Y.~Yu, ``{SIFT}: Building an internet of safe things,'' in
  \emph{Proceedings of the 14th International Conference on Information
  Processing in Sensor Networks}.\hskip 1em plus 0.5em minus 0.4em\relax ACM,
  2015, pp. 298--309.

\bibitem{Raza20132661}
S.~Raza, L.~Wallgren, and T.~Voigt, ``{SVELTE}: Real-time intrusion detection
  in the internet of things,'' \emph{Ad Hoc Networks}, vol.~11, no.~8, pp. 2661
  -- 2674, 2013.

\bibitem{Gisdakis:2015:SDV}
S.~Gisdakis, T.~Giannetsos, and P.~Papadimitratos, ``{SHIELD}: A data
  verification framework for participatory sensing systems,'' in
  \emph{Proceedings of the 8th ACM Conference on Security \& Privacy in
  Wireless and Mobile Networks}, ser. WiSec '15.\hskip 1em plus 0.5em minus
  0.4em\relax ACM, 2015, pp. 16:1--16:12.

\bibitem{F-Secure:2016:sense}
\BIBentryALTinterwordspacing
{F-Secure SENSE}. [Online]. Available: \url{https://sense.f-secure.com/}
\BIBentrySTDinterwordspacing

\bibitem{F-Secure:2016:sense-features}
\BIBentryALTinterwordspacing
What are the current protection features for f-secure sense? [Online].
  Available:
  \url{https://community.f-secure.com/t5/F-Secure-SENSE/What-are-the-current-protection/ta-p/82972}
\BIBentrySTDinterwordspacing

\bibitem{kurtz2016fingerpritning}
A.~Kurtz, H.~Gascon, T.~Becker, K.~Rieck, and F.~Freiling, ``Fingerprinting
  mobile devices using personalized configurations,'' \emph{Proceedings on
  Privacy Enhancing Technologies}, vol.~1, pp. 4--19, 2016.

\bibitem{bojinov2014mobile}
H.~Bojinov, Y.~Michalevsky, G.~Nakibly, and D.~Boneh, ``Mobile device
  identification via sensor fingerprinting,'' \emph{arXiv preprint:1408.1416},
  2014.

\bibitem{Bertini:2015:Profile}
F.~Bertini, R.~Sharma, A.~Iann\`{\i}, and D.~Montesi, ``Profile resolution
  across multilayer networks through smartphone camera fingerprint,'' in
  \emph{Proceedings of the 19th International Database Engineering \&\#38;
  Applications Symposium}, ser. IDEAS '15.\hskip 1em plus 0.5em minus
  0.4em\relax ACM, 2014, pp. 23--32.

\bibitem{VanGoethem2016}
T.~Van~Goethem, W.~Scheepers, D.~Preuveneers, and W.~Joosen,
  ``Accelerometer-based device fingerprinting for multi-factor mobile
  authentication,'' in \emph{Proceedings of the 8th International Symposium on
  Engineering Secure Software and Systems, ESSoS 2016}.\hskip 1em plus 0.5em
  minus 0.4em\relax Springer International Publishing, 2016, pp. 106--121.

\bibitem{Sharaf2016on}
Y.~Sharaf-Dabbagh and W.~Saad, ``On the authentication of devices in the
  {Internet of Things},'' in \emph{Proceedinds of the 17th IEEE International
  Symposium on A World of Wireless, Mobile and Multimedia Networks
  (WoWMoM)}.\hskip 1em plus 0.5em minus 0.4em\relax IEEE, 2016, pp. 1--3.

\bibitem{Haider:2016:Trusted}
I.~Haider, M.~H\"{o}berl, and B.~Rinner, ``Trusted sensors for participatory
  sensing and iot applications based on physically unclonable functions,'' in
  \emph{Proceedings of the 2Nd ACM International Workshop on IoT Privacy,
  Trust, and Security}, ser. IoTPTS '16.\hskip 1em plus 0.5em minus 0.4em\relax
  ACM, 2016, pp. 14--21.

\bibitem{WiFiAlliance2014}
{WiFi Alliance}, \emph{{WiFi} Simple Configuration Technical Specification}.

\end{thebibliography}


\end{document}